\documentclass[journal]{IEEEtran}
 \usepackage[dvips]{graphicx}
   \usepackage{psfrag}
 \DeclareGraphicsExtensions{.eps}

\usepackage{array}
\usepackage{cite}
\usepackage{amsmath}
\usepackage[absolute]{textpos}
\DeclareMathOperator*{\argmax}{arg\,max}
\usepackage{amssymb}
\usepackage[linesnumbered,lined,boxed,commentsnumbered]{algorithm2e}
\usepackage[usenames, dvipsnames]{color}

\usepackage{subfigure}

\usepackage{url}

\newcommand{\copyrightstatement}{
    \begin{textblock}{15}(0.5,0.3)    % tweak here: {box width}(leftposition, rightposition)
         \noindent
         \centering
         \textblockcolour{white}
         \footnotesize
         \copyright 2017 IEEE. Personal use of this material is permitted. Permission from IEEE must be obtained for all other uses, in any current or future media, including reprinting/republishing this material for advertising or promotional purposes, creating new collective works, for resale or redistribution to servers or lists, or reuse of any copyrighted component of this work in other works
    \end{textblock}
}
\begin{document}
\title{Decoding-Energy-Rate-Distortion Optimization \\ for Video Coding}
\copyrightstatement

\author{Christian~Herglotz,
        Andreas~Heindel,
        and~Andr\'e~Kaup,~\IEEEmembership{Fellow,~IEEE}% <-this % stops a space
        \thanks{Copyright \copyright ~2017 IEEE. Personal use of this material is permitted. However, permission to use this material for any other purposes must be obtained from the IEEE by sending an email to pubs-permissions@ieee.org.}
\thanks{This work was financially supported by the Research Training Group 1773 ``Heterogeneous Image Systems'', funded by the German Research Foundation (DFG).}
\thanks{C. Herglotz, A. Heindel, and A. Kaup are with the Chair of Multimedia Communications and Signal Processing, Friedrich-Alexander-Universit\"at Erlangen-N\"urnberg (FAU), Erlangen, Germany (\{ christian.herglotz, andreas.heindel, andre.kaup \} @FAU.de). }% <-this % stops a space
%\thanks{Manuscript received September 1, 2014; revised December 27, 2014.}
}

% make the title area
\maketitle

% As a general rule, do not put math, special symbols or citations
% in the abstract or keywords.
\begin{abstract}
This paper presents a method for generating coded video bit streams requiring less decoding energy than conventionally coded bit streams. To this end, we propose extending the standard rate-distortion optimization approach to also consider the decoding energy. In the encoder, the decoding energy is estimated during runtime using a feature-based energy model. These energy estimates are then used to calculate decoding-energy-rate-distortion costs that are minimized by the encoder. This ultimately leads to optimal trade-offs between these three parameters. % decoding energy, rate, and distortion. 
Therefore, we introduce the mathematical theory for describing decoding-energy-rate-distortion optimization and the proposed encoder algorithm is explained in detail. For rate-energy control, a new encoder parameter is introduced. Finally, measurements of the software decoding process for HEVC-coded bit streams are performed. Results show that this approach can lead to up to \textbf{$30\%$} of decoding energy reduction at a constant visual objective quality when accepting a bitrate increase at the same order of magnitude. 
\end{abstract} 

% Note that keywords are not normally used for peerreview papers.

\section{Introduction}
\IEEEPARstart{I}{n} recent years, the global demand for video streaming services %(e.g., YouTube or Netflix) 
has increased rapidly. Nowadays, video data constitutes more than $50\%$ of the mobile data traffic \cite{cisco17}. In order to optimally exploit the available wireless channel capacity, all video coding standards such as Advanced Video Coding (H.264/AVC) \cite{ITU_H.264} or the most recent High Efficiency Video Coding (HEVC) standard \cite{ITU_HEVC} primarily aim at reducing the bitrate while maintaining the objective visual quality of the sequence. Simultaneously, a dramatic increase in encoding and decoding complexity due to more sophisticated and efficient compression algorithms has been observed. An analysis of the power consumption of the decoding process for a H.264 software decoder on a smartphone was performed in \cite{Carroll13} and showed that the major part of the device's power consumption is used for processing on the central processing unit (CPU). Hence, research aiming at reducing the decoder's processing power is of great interest. 

To this end, we propose modifying the classic rate-distortion optimization (RDO) approach \cite{Sullivan98} used in many encoder solutions such as the HEVC Test Model (HM) reference software for the HEVC codec \cite{HM} or the JM reference software for the H.264/AVC codec \cite{JM}. The goal of the state-of-the-art RDO is to minimize the distortion $D$ assuming the constraint that the rate $R$ does not exceed the maximum value $R_\mathrm{max}$. Mathematically, this optimization problem can be formulated as 
\begin{align}
\min D \quad \mathrm{s.t.} \quad  R \le R_\mathrm{max}. 
\label{eq:RD_base}
\end{align}
As this representation is inconvenient for practical use, an equivalent formula was derived in \cite{Shoham88} using a Lagrange multiplier $\lambda$. The formula then reads 
\begin{equation}
\min J = D+\lambda R, 
\label{eq:RD_lagr}
\end{equation}
where $J$ represents the costs to be minimized. Here, $\lambda$ can be interpreted as the trade-off parameter between distortion and rate. In a typical video codec implementation such as JM or HM, the cost $J$ is calculated for each coding mode considered and the mode exhibiting the lowest cost is chosen. 

In this paper, we propose extending \eqref{eq:RD_lagr} to decoding-energy-rate-distortion optimization (DERDO) and propose including the decoding energy $E$ into the formula as 
\begin{equation}
\min J = D+\lambda_R\cdot R + \lambda_E \cdot E, 
\label{eq:RD_lagrE}
\end{equation}
where we incorporate a second Lagrange multiplier $\lambda_E$ that, in conjunction with $\lambda_R$, determines the trade-off between distortion, rate, and decoding energy. Using this optimization function, the encoder is able to construct bit streams that 
\begin{itemize}
\item comply with the standard and
\item require less processing energy during the decoding process. 
\end{itemize}
%Hence, using this approach, the decoding energy on any device can be minimized without having to reconfigure the software or the hardware. 
Hence, using this approach, the decoding energy can be minimized without having to reconfigure the software or the hardware of the decoding device. The only prerequisite is that the processing energy depends on the use of coding tools, which is given for any software decoder. Please note that in this work, evaluations are only performed for software decoders showing that DERDO is applicable in practice. For hardware decoders, this approach is potentially applicable if, e.g., various components like an in-loop filtering stage can be switched on and off during decoder runtime. Such a decoder is, e.g., the hardware chip proposed by Engelhardt et al. \cite{Engelhardt14}. %However, the generalization to hardware decoders is an interesting topic for future work. 

For the HEVC intraframe coding case using a software decoder, it was shown that around $15\%$ of decoding energy could be saved when accepting a bitrate increase of around $15\%$ \cite{Herglotz16a}. The drawback of a higher bitrate can be accepted when the available bandwidth is sufficient, which is true in many real applications. For example, our tests showed that a class C sequence ($832\times 480$ pixels) encoded with HEVC at a medium quality (quantization parameter (QP) equal to $30$) requires a bandwidth of about $2.4$ Mbit/s which is much lower than the typical bandwidth provided by modern WiFi standards such as 802.11n, which increases to nearly $100$ Mbit/s \cite{Sun14}. 
% I was using BasketballDrill QP 32

A major challenge imposed by this approach is determining the decoding energy $E$ during runtime in the encoder. As a solution, we propose making use of a feature based energy model as presented in \cite{Herglotz16c} where it was shown that such an energy model could be constructed for any hybrid video codec. In this model, each feature can be described by a certain standard process in the decoder, e.g., a feature can correspond to the residual transformation of a certain block size. As the corresponding reconstruction process of this block has a defined and homogeneous processing flow with a rather constant amount of processing energy for each execution, the corresponding processing energy can be derived using a dedicated measurement and training method. The obtained information can then be used inside the encoder to decide on the coding mode that requires a small amount of energy. 

The paper is organized as follows: First, Section \ref{sec:lit} reviews relevant literature and puts our approach into the scientific context. Afterwards, Section \ref{sec:model} introduces the various aspects of energy consumption during video decoding and presents respective energy models. %In Section \ref{sec:training}, the different aspects of model parameter training are elucidated. 
Subsequently, Section \ref{sec:DERDO} discusses the theory behind DERDO and explains how DERDO can be included in a given encoder. Afterwards, Section \ref{sec:eval} applies DERDO to the HEVC encoder and discusses its theoretical and actual performance, as obtained by real measurements. Finally, conclusions are given in Section \ref{sec:concl}.

\section{Literature Review}
\label{sec:lit}

A great deal of scientific contributions can be found regarding energy efficient video coding. A major subset of these works aims at optimizing implementations in hardware or software. Examples for specialized hardware can be found for intraframe coding \cite{Kalali12}, transformation \cite{Tikekar14}, arithmetic decoding \cite{Chen14b}, HEVC's sample adaptive offset (SAO) filter \cite{Park13}, and many more. Other works, e.g. \cite{Li12DRAM}, focus on the power consumption of dynamic random access memory (DRAM) where smart data manipulation techniques can reduce the power consumption by more than $90\%$. Furthermore, several software encoders and decoders are implemented and continuously refined for real-time streaming purposes \cite{FFmpeg,libde}. 

Another branch focuses on the optimization of the encoder. In the commonly used rate-distortion (RD) approach \cite{Sullivan98}, nearly all possible coding modes are tested for rate and distortion and the best mode is chosen for coding. This brute-force search shows a very high computational complexity such that current research tries to enable real-time encoding while losing little performance in the RD sense. For H.264/AVC, such an approach is proposed in \cite{Li11} where a corresponding complexity metric is derived for the testing of each coding mode. Then, depending on the available computational power, only a part of the possible coding modes is tested for RD costs such that the encoding time can be reduced significantly. A similar approach was presented in \cite{He05} for a power-saving H.263 encoder. Furthermore, subsequent work included other energy reduction techniques such as Dynamic Voltage and Frequency Scaling (DVFS) \cite{Liang09}.

On the decoder side, mainly four approaches can be found in the literature. In the first approach, similar to the approach used for the encoder, DVFS can be used for saving energy. An early work for H.263 was proposed by Pouwelse et al. \cite{Pouwelse01} in which the decoding device switches between different voltage and frequency states depending on the frame type (I, P, or PB-frame). A more sophisticated solution was presented in \cite{Landge05} for a 3D Wavelet based Scalable Codec (SVC) \cite{Ohm04}.  It was proposed to transmit further side information from the encoder side using the MPEG-21 standard. In the second approach, Pakdeepaiboonpo et al. proposed optimizing the compiler for an MPEG-4 decoder \cite{Pakdeepaiboonpol05} and achieved energy savings of up to $13\%$. 

The third approach focuses on the type of the bit stream. In that regard, Ren et al. \cite{Ren14b} proposed switching between different coding standards. When the battery is low, the decoder can switch from HEVC decoding to H.264/AVC decoding and extend the operating time by more than $5\%$. 

In the last approach, which is similar to our proposal, the decoding power or complexity is estimated on the encoder side such that energy saving bit streams are constructed. Van der Schaar et al. \cite{vdSchaar05} profiled a decoder for the spatial-domain motion-compensated temporal filtering (SDMCTF) decoder \cite{Andreopoulos04} and built a generic complexity model based on processor operations. Lee et al. \cite{Lee07b} constructed a similar model for H.264/AVC but focused on motion compensation. Finally, He et al. \cite{He13} modeled motion compensation and the in-loop filter in HEVC and proposed switching to energy saving bit streams when the battery is low. 

In contrast to the aforementioned work, we do not base our model on complexity metrics but on actual energy measurements. Furthermore, our modeling approach is more general as we aim at modeling the complete decoding process (see Section \ref{sec:model}). In contrast, other work only focused on selected parts of the decoder. Basically, we are capable of estimating the decoding energy for each possible coding mode with a high precision such that significant energy savings can be achieved. %This information is used to perform an elaborate analysis of the encoder in terms of decoding-energy-rate-distortion (DERD) performance. 

\section{Energy Consumption of Video Decoders}
\label{sec:model}
Determining and modeling the processing energy of a modern device is a highly complex task. To this end, Carrol et al. \cite{Carroll13} performed a general analysis of the power and energy consumed by a portable smartphone. The analysis was conducted for a high-end smartphone that was released in 2012. As the major setup of a portable device did not change significantly, the results are still valid for modern devices. 

In their analysis, a power breakdown for software decoding of a high and a low-quality H.264-coded bit stream was determined. A relevant workload was found on the following components of the smartphone: RAM, graphics processing unit (GPU), internal hardware systems, core, memory interface (MIF), display, and remaining miscellany of the system on chip (SoC). Except for the display, all of the aforementioned components are also considered in our model (including the CPU and the RAM). %Corresponding measurements were conducted for the energy consumption of the complete device \cite{Herglotz16b}.
 The display is not considered as its power consumption is mainly determined by the backlight brightness \cite{Carroll13} which cannot be controlled by our approach. A more detailed explanation on the measurement setup is given in Section \ref{secsec:meas}.

Note that in this work, we do not consider the transmission energy when using WiFi or a GSM network, for reasons of brevity. In \cite{Herglotz16a}, the influence of the transmission energy in a WiFi-streaming scenario for HEVC-intra coding was tested. The energy savings were found to be smaller, but only by less than $5\%$. Since the rates and, hence, transmission energies are much lower for the general inter-coding case we address here, we can say that this simplification is valid. Further work should examine this problem more closely. 

We estimate the energy consumption during video decoding $\hat E_\mathrm{dec}$ that is mainly composed of the CPU and RAM energy. Various models have been proposed to estimate this energy. For the HEVC decoder, a helpful overview is given in \cite{Herglotz16b}, where a high amount of different models was introduced and compared in terms of estimation accuracy. For our work, we concentrate on two types of models: A \textit{high-level model} (HL) and a \textit{bit stream feature based model} (BF). The former model provides insights into theoretical DERD considerations (see Section \ref{secsec:lambdaQP}) while the latter is used for explicit DERDO (Section \ref{secsec:impl}) in the encoder. In \cite{Herglotz16c} we showed that both models can be employed for various codecs such that the considerations given in this section apply generally. 

For the \textit{high-level model}, the decoding energy is estimated using high-level parameters, namely the resolution in pixels per frame $S$, the number of frames $N$, the bit stream file size $B$, and the fraction of intra coded frames $p_\mathrm{I}$. $p_\mathrm{I}$ is the share of intra coded frames of all frames ($p_\mathrm{I}=\frac{N_\mathrm{I}}{N}$ with the number of I-frames $N_\mathrm{I}$).  We consider a simple linear model that was originally proposed in \cite{Raoufi13} and that targets the frame-based decoding energy. In \cite{Herglotz16b}, it was generalized to the decoding energy of a complete sequence. It is given by % and reads
\begin{equation}
\hat E_\mathrm{dec,HL} = \left( c_1 \cdot p_\mathrm{I} \cdot \frac{B}{N\cdot S} + c_2 \cdot p_\mathrm{I} + c_3 \cdot \frac{B}{N\cdot S} + c_4\right) \cdot N \cdot S. 
\label{eq:Raoufi}
\end{equation}
In essence, this model estimates the mean per-pixel decoding energy based on the mean number of coded bytes per pixel. Using the parameter $p_\mathrm{I}$, the difference between the decoding energy for inter and intra coded frames can be considered. 

The parameters can be interpreted as follows: The constant $c_4$ considers a minimum energy required to decode one pixel. $c_2$ can be interpreted as an additional energy required when the pixel is located inside an I-frame. $c_3$ describes the additional energy depending on the bitrate: The assumption is that the energy rises linearly with the mean number of coded bits per pixel. Finally, $c_1$ is used in a mixed term referring to both the mean bits per pixel and the fraction of I-frames. Please note that the latter two constants %, which link the energy with the bitrate, 
are used to develop the DERD theory in Section \ref{secsec:lambdaQP}. Due to the empirical relation between the bit stream file size and the energy, incorporating the energetic properties into the analytic rate-distortion function is possible. %s, as shown in Section \ref{sec:DERDO}. 

These constants $c_1,...,c_4$ describe the energetic properties of the decoding system. For sufficiently long sequences \cite{Herglotz16b} and when the encoder settings and the type of input sequence are fixed \cite{Herglotz16c}, the estimation error is reported to be smaller than $22\%$ for all tested codecs and decoder implementations. 

%In this work, the HL-model is exploited to discuss DERD theory. 

The \textit{bit stream feature based model} can be described by a linear equation: 
\begin{equation}
\hat E_\mathrm{dec,BF} = \sum_{ f \in F} n_f \cdot e_f. 
\label{eq:BFM}
\end{equation}
In this model, a constant set $F$ of bit stream features $f$ is defined. The number of occurrences of each of these features in a coded bit stream is represented by the feature number $n_f$. Each of these features is described by a corresponding specific energy $e_f$ that can be interpreted as the processing energy consumed each time this feature occurs. 

To provide a vivid example, one feature can be the transformation of a residual block at a fixed block size. Interpreting the bit stream we can count how often this process must be executed for reconstructing the complete sequence. As the transformation itself has a rather constant processing flow for a given block size, the corresponding processing energy on the CPU is nearly constant and can be determined using a suitable training method. Generally, the same holds for the RAM energy because the amount of memory accesses (read and write) is also nearly constant for each execution. 

Please note that considering the RAM energy, at the beginning of the decoding process, the data is not yet available in the cache such that the RAM is accessed more often resulting in a significantly higher energy. Hence, for single frame energy estimation, the estimation errors are relatively high (more than $15\%$ \cite{Herglotz16b}). Furthermore, other processes that are running concurrently and that replace data in the cache can lead to energy variations and bad estimations. However, in \cite{Herglotz16b} it was shown that for different processors (ARM and Intel), the RAM energy can be estimated accurately (mean estimation error below $5\%$) for sufficiently long sequences (more than $8$ frames which is given for most practical applications). Hence, the model is generally applicable for both the CPU and the RAM energy consumption of video decoders. 

The set of features used in this work corresponds to the simple model presented in detail in \cite{Herglotz16b}. A list summarizing these features is given in Table \ref{tab:featList}. 
\begin{table}[t]
\caption{Features used in the bit stream feature based model ($27$ features). The depth indicates the quadtree depth, if applicable. A depth of $0$ corresponds to a block size of $64\times 64$ and goes down to a block size of $4\times 4$ at depth $4$. A more detailed explanation can be found in \cite{Herglotz16b}.   }
\label{tab:featList}
\begin{center}
\begin{tabular}{l|l|c}
\hline
Label & Depths & Explanation \\
\hline
offset & - & constant offset \\
I\_frames & -  & number of I frames \\
PB\_frames & - & number of P and B frames\\
intraCUs & - & complete number of intra CUs\\
intraPUs& $1..4$ & intra predicted blocks\\
skip& $0..3$& skipped blocks\\
inter& $0..3$ & inter and merge blocks\\
fracpel& - & fractional pel filterings\\
coeffs& - & nonzero coefficients\\
val& - & summed logarithmic \\
& & values of nonzero coefficients\\
trans& $1..4$ & transformations \\
bi& $4$ & bi-predicted $4\times 4$ blocks\\
Bs& - & boundaries with deblocking filtering\\
SAO\_Y& $0$ & SAO-filtered luma CTUs\\
SAO\_C& $0$ & SAO-filtered chroma CTUs\\
\hline
\end{tabular}
\end{center}
\end{table}
In terms of the combined energy consumption (RAM and CPU), in \cite{Herglotz16b} it was shown that the bit stream feature based model achieves the lowest estimation errors for HEVC ($<10\%$) in comparison to other tested models. Furthermore, in \cite{Herglotz16c} it was shown that the same observation holds for different codec's decoders. Note that this model enables counting all features during encoding and hence, calculating decoding energy estimates. Thus, this kind of model is well suited for explicit DERDO in the encoder as will be discussed in Section \ref{secsec:impl}.

\section{Decoding-Energy-Rate-Distortion Optimization}
\label{sec:DERDO}

Being able to accurately estimate the potential decoding energy during the encoding process, we can incorporate the energy into the classic rate-distortion optimization algorithm. Therefore, in the first subsection, we formally define the new optimization function. Afterwards, we discuss the theoretical impact of the decoding energy on the solution of the optimization problem and derive a new relation between the Lagrange multipliers and the quantization parameter (QP).

\subsection{Formal Derivation}
Similar to (\ref{eq:RD_base}), we define a maximum desired rate $R_\mathrm{max}$ to limit the channel throughput and a maximum desired decoding energy $E_\mathrm{max}$ depending on the battery charging, amongst others. Given these constraints we attempt to minimize the distortion $D(s)$ and obtain the optimization formula
\begin{eqnarray}
\min_{s\in S} J(s) =  D(s) &  \mathrm{s.t.} &  R(s) \le R_\text{max}  \label{eq:minJ_RD_base} \\
&	& E(s) \le E_\text{max},   \notag
\end{eqnarray}
where $s$ is one feasible solution for encoding from the complete set of feasible solutions $S$. This formulation is problematic as usually a maximum rate and energy cannot be guaranteed during encoder runtime as they greatly depend on the content of the sequence. As a solution, a Lagrange relaxation is usually employed to make sure that the chosen solution is optimal for an arbitrary maximum rate and energy. We maintain this approach and, similarly to the proof given in \cite{Shoham88}, prove that 
\begin{equation}
\min J(s) = D(s)+\lambda_R R(s) + \lambda_E E(s)
\label{eq:minJ_lambda}
\end{equation}
is an equivalent representation. 

Consider $\lambda_R$ and $\lambda_E$ are greater than zero. Then we can find a solution $s^*(\lambda_R, \lambda_E)$ such that (\ref{eq:minJ_lambda}) is minimal. Now we set $R_\mathrm{max} = R(s^*)$ and $E_\mathrm{max} = E(s^*)$. Due to the minimization, the following inequation holds 
\begin{align}
 D(s^*)+\lambda_R\cdot R(s^*)+\lambda_E\cdot E(s^*) &  \notag \\ 
 \le D(s)+\lambda_R\cdot R(s)+\lambda_E\cdot E(s), & \ \forall s\in S \label{eq:s*lower}
\end{align}
that can be rewritten as 
\begin{align}
D(s^*)-D(s) & \notag \\
\le \lambda_R\left[R(s)-R(s^*)\right] + \lambda_E\left[E(s)-E(s^*)\right], & \ \forall s\in S. \label{eq:s*lower2}
\end{align}
Now we define a subset $\bar S \subset S = \{ s:R(s)\le R(s^*)\} \cap \{s:E(s)\le E(s^*)\}$ containing all solutions that show a smaller or equal rate and a smaller or equal energy, i.e. a subset that fulfills the constraints given in (\ref{eq:minJ_RD_base}). For this subset, $\left[R(s)-R(s^*)\right]\le 0$ and $\left[E(s)-E(s^*)\right]\le 0$. Consequently, as $\lambda_R$ and $\lambda_E$ are positive by definition, the right side of (\ref{eq:s*lower2}) is smaller than or equal to zero and hence 
\begin{equation}
D(s^*)-D(s)\le 0 \Rightarrow D(s^*) \le D(s), \ \forall s\in \bar S, 
\label{eq:s*lowerproof}
\end{equation}
which means that for all solutions that satisfy the constraints $s^*$ is the optimal one.

\subsection{$\lambda$-QP Relation}
\label{secsec:lambdaQP} 
One of the major reasons for the encoder's complexity is the possibility to choose between different QP values. For example, in HEVC, the QP ranges from $0$ to $51$ and theoretically, for a given fixed $\lambda_R$, every possible QP value must be checked for RD performance to find the best solution. As a relief, assuming that the Gaussian rate-distortion function and the high-rate approximation hold \cite{Sullivan98}, an analytic relation between $\lambda_R$ and its corresponding optimal and constant QP can be derived. Using this approach, usually only a single QP for a given $\lambda_R$ is tested in current encoder implementations such that encoding times are significantly lower. Several empirical studies proved that this approach only leads to minor RD losses (cf., \cite{Sullivan98, binli13}). 

In this section, we discuss the theoretical implication of such a simplification and its feasibility for DERDO. 
To this end, we rewrite the optimization formula (\ref{eq:RD_lagrE}) such that it only depends on the distortion $D$ as 
\begin{equation}
\min J(D)  =  D + \lambda_R R(D) + \lambda_E E(R(D)), 
\label{eq:J_E_insert}
\end{equation}
where the energy, in accordance with the \textit{high-level energy model} (\ref{eq:Raoufimod}), is interpreted as a function of the rate, which again is a function of the distortion. 
Calculating the derivative with respect to $D$, as indicated by the prime, the local minima can be computed as 
\begin{equation}
\frac{dJ}{dD} = 0 = 1+\lambda_R \cdot R'(D) + \lambda_E \cdot E'(R(D))\cdot R'(D). 
\label{eq:J_derived}
\end{equation}
When this equation is solved for $D$, we can calculate a corresponding quantization step size $q$ using the common high-rate approximation \cite{Gish68}
\begin{equation}
D \approx \frac{\left(2q\right)^2}{12}. 
\label{eq:highRateAss}
\end{equation}

The formulation shown in (\ref{eq:J_derived}) can be analyzed when an analytic relation between distortion and rate as well as rate and energy is given. For the $R(D)$ relation, we adopt the common rate-distortion function for a memoryless Gaussian source \cite{Berger71}
\begin{align}
R(D) = & \frac{1}{2}\max\left(0, \log\left( \frac{\sigma^2}{D}\right)\right), %\notag \\
\label{eq:gaussRD} 
\end{align}
where $\sigma^2$ is the variance of the source signal and $\log$ corresponds to the natural logarithm. % to the base $e$. 
Considering the case $R>0$ the derivative is given by 
\begin{equation}
R'(D) = -\frac{1}{2}\frac{D}{\sigma^2}\frac{\sigma^2}{D^2} = -\frac{1}{2D}, 
\label{eq:R'(D)}
\end{equation}
which is independent from $\sigma^2$. 

For the relation between rate and decoding energy $E(R)$, we consider the \textit{high-level model} (cf. Section \ref{sec:model}) as it gives valuable insights into the optimization problem. To maintain general applicability, observations will be performed independent from the resolution $S$ and the length $N$ by generalizing to energy per pixel and rate per pixel as 
\begin{equation}
\hat E_\mathrm{dec,pixel} = \frac{\hat E_ \mathrm{dec}}{N\cdot S}, \quad R = \frac{B}{N\cdot S} 
\label{eq:perPixelVals}
\end{equation}
which we insert into the model's estimation function 
(\ref{eq:Raoufi}) and obtain 
\begin{equation}
\hat E_\mathrm{dec,pixel}(R) = c_1 \cdot p_\mathrm{I} \cdot R + c_2 \cdot p_\mathrm{I} + c_3 \cdot R + c_4. 
\label{eq:Raoufimod}
\end{equation}
The corresponding derivative yields
\begin{equation}
\hat E_\mathrm{dec,pixel}'(R) =  c_1 \cdot p_\mathrm{I}+ c_3 . 
\label{eq:Raoufimod'}
\end{equation}
Hence, inserting (\ref{eq:Raoufimod'}) and (\ref{eq:R'(D)}) into (\ref{eq:J_derived}) we obtain 
\begin{equation}
\frac{dJ}{dD} = 0 = 1- \frac{\lambda_R}{2D} - \left(  c_1 \cdot p_\mathrm{I}+ c_3   \right) \cdot \frac{\lambda_E}{2D}
\label{eq:J_derived_insert_Raoufi}
\end{equation}
which we can solve for $D$ and insert into (\ref{eq:highRateAss}), resulting in an analytic expression linking the quantization step size to the Lagrange multipliers as 
\begin{equation}
q = \frac{1}{2} \sqrt{12\left[ \frac{\lambda_\mathrm{R}}{2}+\frac{\lambda_\mathrm{E}}{2}\left(c_1 \cdot p_\mathrm{I}+ c_3   \right) \right]}. 
\label{eq:lambdaQP_Raoufi}
\end{equation}
%To obtain the final $\lambda$-QP relation
For newer standards such as HEVC or H.264/AVC, we additionally insert the relation between $q$ and QP \cite{Budagavi13} as 
\begin{equation}
q=2^{\frac{\mathrm{QP}-4}{6}},  
\label{eq:qQP}
\end{equation}
which is valid for $8$ bit sequences (the equation for $10$ bit sequences can be applied similarly). We obtain 
\begin{equation}
\mathrm{QP} = 4+6\cdot \mathrm{ld}\left( \sqrt{\frac{3}{2}\lambda_\mathrm{R} + \frac{3}{2}\lambda_\mathrm{E}\left( c_1 \cdot p_\mathrm{I}+ c_3\right)} \right), 
\end{equation}
where ld corresponds to the logarithm to the basis $2$. Finally, we eliminate the square root and define two new variables 
\begin{equation}
\rho = \frac{3}{2}, \qquad \varepsilon = \frac{3}{2}\left( c_1 \cdot p_\mathrm{I}+ c_3\right)
\label{eq:rhoepsilon}
\end{equation}
 to summarize the factors of the Lagrange multipliers. Please note that $\varepsilon$ depends on the parameters $c_1$ and $c_3$ of the \textit{high-level model} that link the bitrate with the decoding energy (cf. Section \ref{sec:model} for more details on the parameters). Hence, their value and thus the value of $\varepsilon$ depends on the employed decoding system. With these parameters, we obtain the general expression
\begin{equation}
\mathrm{QP} = 4+3\cdot \mathrm{ld}\left( \rho \lambda_\mathrm{R} + \varepsilon\lambda_\mathrm{E}\right)
\label{eq:QP_lambda_final}
\end{equation}
which links the two Lagrange multipliers with the QP. For other standards using the quantization step size $q$, the corresponding general expression reads
\begin{equation}
q = \sqrt{\rho \cdot \lambda_\mathrm{R}+\varepsilon\cdot\lambda_\mathrm{E}}. 
\label{eq:lambdaQP_general}
\end{equation}

\subsection{Rate-Energy Control}
\label{secsec:r_e_ctrl}
In the next step, the $\lambda$-values must be determined. In classic RDO (e.g. the HM reference software \cite{HM} for the HEVC standard), $\lambda$ is often calculated using the QP that is selected by the user. This approach has the following advantages: 
\begin{itemize}
\item The user can set the desired visual quality in terms of the quantization step size, 
\item Its range of values is much smaller (0...51 in comparison to $0.01$...$10000$) and hence more convenient, 
\item The $\lambda$-value itself is irrelevant for decoding (it is not part of the standard) such that it can be used as an internal variable that users do not have to consider. 
\end{itemize}
Unfortunately, in our new approach using two $\lambda$s, a single parameter controlling the $\lambda$-values is no longer sufficient. Therefore, we propose using a new user-defined parameter $\tau$ that controls the rate-energy trade-off. For convenience and ease of use it should have the following properties: 
\begin{itemize}
\item The classic QP should still be used, 
\item Its value should be easy to interpret, 
\item The parameter should cover the range from pure RDO to pure DEDO, 
\item Classic evaluation methods such as the well-known Bj{\o}ntegaard-Delta \cite{Bjonte01} should be easily applicable. 
\end{itemize}
Hence, we define the parameter to be situated in the range $\tau \in [0;1]$, where $\tau =0$ means pure RDO that corresponds to $\lambda_\mathrm{E}=0$, $\tau =0.5$ means equal ``importance'' for rate and energy, and $\tau =1$ means pure decoding energy optimization ($\lambda_\mathrm{R}=0$). 

With these conditions in mind we look for a suitable function that maps this range to the potential range of $\lambda$-values. Therefore, we consider the relation of the two summands in (\ref{eq:QP_lambda_final}) $\frac{\rho\lambda_\mathrm{R}}{\varepsilon\lambda_\mathrm{E}}\in [0;\infty)$ which is positive since both summands are always positive. A suitable function that maps this range to the interval $[0;1]$ is given by 
\begin{equation}
\frac{\rho\lambda_\mathrm{R}}{\varepsilon\lambda_\mathrm{E}} = \left( \frac{1}{\tau}-1\right)^a, 
\label{eq:tau}
\end{equation}
where we choose $a=3$ heuristically as it enables a quasi-linear trade-off between rate and decoding energy performance (cf. RD and DED curves in Section \ref{sec:eval}). 

Using (\ref{eq:QP_lambda_final}), \eqref{eq:lambdaQP_general}, and \eqref{eq:tau} we now obtain analytic functions that can be used to calculate the two Lagrange multipliers by 
\begin{equation}
\lambda_\mathrm{R} = \frac{q^2}{\rho\cdot\left( 1+ \left(\frac{1}{\tau}-1\right)^{-a}\right)} = 
	\frac{2^{\frac{\mathrm{QP}-4}{3}}}{\rho\cdot\left( 1+ \left(\frac{1}{\tau}-1\right)^{-a}\right)}
\label{eq:lambdaRofQPtau}
\end{equation}
and 
\begin{equation}
\lambda_\mathrm{E} = \frac{q^2}{\varepsilon\cdot\left( 1+ \left(\frac{1}{\tau}-1\right)^{a}\right)}
				= \frac{2^{\frac{\mathrm{QP}-4}{3}}}{\varepsilon\cdot\left( 1+ \left(\frac{1}{\tau}-1\right)^{a}\right)},  
\label{eq:lambdaEofQPtau}
\end{equation}
where the left fractions are used for the quantization $q$ and the right fractions for the QP ($8$ bit sequences).

\begin{figure}
\centering
\psfrag{000}[c][c]{$0$}
\psfrag{001}[c][c]{$0.5$}
\psfrag{002}[c][c]{$1$}
\psfrag{003}[c][c]{$0$}
\psfrag{004}[c][c]{$20$}
\psfrag{005}[c][c]{$40$}
\psfrag{006}[c][b]{$60$}
\psfrag{007}[r][r]{$10^{-5}$}
\psfrag{008}[r][r]{$1$}
\psfrag{009}[r][r]{$10^5$}
\psfrag{010}[c][c]{$\lambda_\mathrm{R}$}
\psfrag{011}[c][c]{QP}
\psfrag{012}[c][c]{$\tau$}
\includegraphics[width=0.48\textwidth]{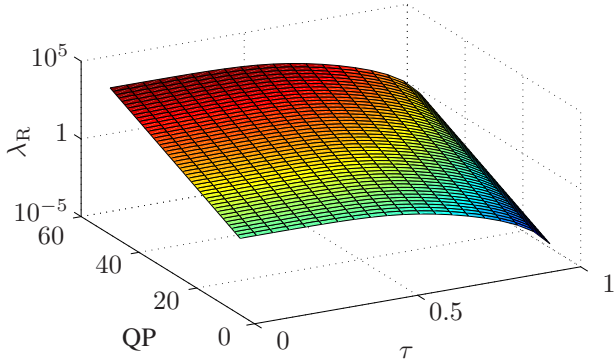}
\caption{Analytic relation between the Lagrange multiplier $\lambda_\mathrm{R}$, the QP, and $\tau$ using (\ref{eq:lambdaRofQPtau}). $\rho$ is set to $\frac{3}{2}$ and $\varepsilon$ is calculated with $p_\mathrm{I}=0.02$, $c_1=-2.9\cdot 10^{-7}$, and $c_3=9.7\cdot 10^{-7}$. The latter two values were fitted for the main evaluation setup shown in Section \ref{secsec:meas}.  }
\label{fig:lambdaQP_lR}
\end{figure}

\begin{figure}
\centering
\psfrag{000}[c][c]{$0$}
\psfrag{001}[c][c]{$0.5$}
\psfrag{002}[c][c]{$1$}
\psfrag{003}[c][c]{$0$}
\psfrag{004}[c][c]{$20$}
\psfrag{005}[c][c]{$40$}
\psfrag{006}[c][b]{$60$}
\psfrag{007}[r][r]{$1$}
\psfrag{008}[r][r]{$10^5$}
\psfrag{009}[r][r]{$10^{10}$}
\psfrag{010}[r][r]{$10^{15}$}
\psfrag{011}[c][c]{$\lambda_\mathrm{E}$}
\psfrag{012}[c][c]{QP}
\psfrag{013}[c][c]{$\tau$}
\includegraphics[width=0.48\textwidth]{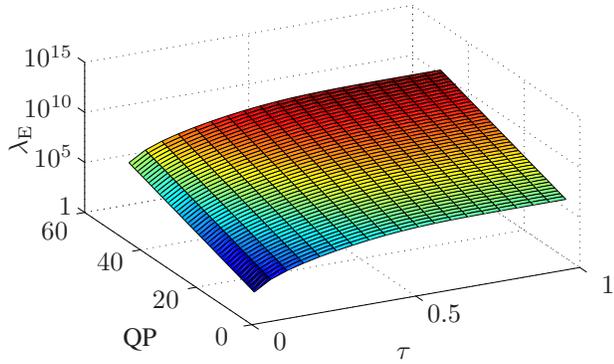}
\caption{Analytic relation between the Lagrange multiplier $\lambda_\mathrm{E}$, the QP, and $\tau$ using (\ref{eq:lambdaEofQPtau}). For $\rho$ and $\varepsilon$, the values from Fig. \ref{fig:lambdaQP_lR} are used.   }
\label{fig:lambdaQP_lE}
\end{figure}

To visualize the behavior of this curve we calculate $\rho$ and $\varepsilon$ using (\ref{eq:rhoepsilon}) for a special test case. Figures \ref{fig:lambdaQP_lR} and \ref{fig:lambdaQP_lE} show how $\lambda_\mathrm{R}$ and $\lambda_\mathrm{E}$ depend on the choice of the QP and $\tau$, respectively. 
Considering Fig. \ref{fig:lambdaQP_lR}, we can see the behavior of $\lambda_\mathrm{R}$. First, we consider $\tau=0$ which is classic RDO. This case corresponds to the left border of the surface where we can see that $\lambda_\mathrm{R}$ increases logarithmically for an increasing QP as previously shown (cf. \cite{binli13}). If we then increase $\tau$ the values for $\lambda_\mathrm{R}$ decrease. Nevertheless, for every choice of a constant $\tau$ (diagonal lines on the surface), the logarithmic behavior can still be observed. Note that the line for $\tau=1$ is not plotted since $\lambda_\mathrm{R}$ would be 0 for all QPs in this case, which cannot be displayed due to the logarithmic representation of $\lambda_\mathrm{R}$. 

If we now consider Fig. \ref{fig:lambdaQP_lE} we can see that the values for $\lambda_\mathrm{E}$ show an inverse behavior with respect to $\tau$. This means that $\lambda_\mathrm{E}$ increases for increasing values of $\tau$. Note that for a fixed QP, corresponding to \eqref{eq:QP_lambda_final}, the sum of the weighted $\lambda$s (weighted with $\rho$ and $\varepsilon$) is always constant. 

Later, in Section \ref{secsec:evalLam} we show that this theoretical relation correlates well with the results obtained in experiments. Beforehand, in the next section, we explain how DERDO is implemented in a real encoder.

\subsection{DERDO Implementation}
\label{secsec:impl}

To implement the DERDO-cost calculations, we extend an existing encoder solution that makes use of RDO. % we take the HEVC test model software HM-14.0 \cite{HM} as a basis. 
Two major parts must be added: The estimation of the blockwise decoding energy to be used for an optimal DERD decision (Section \ref{secsec:decEnEst}) and the calculation of the new cost function using the estimated energy (Section \ref{secsec:codCostCalc}). With these two modifications, the encoder is able to choose the energetically most efficient coding mode. %The choice of the energetically optimal features is hence based on the modified costs. 

\subsubsection{Encoder-side Decoding Energy Estimation}
\label{secsec:decEnEst}
To estimate the decoding energy $\hat E_\mathrm{dec}$ during encoding, 
we propose using the \textit{bit stream feature-based model} (BF) introduced in Section \ref{sec:model}. 
For demonstration purposes we give an illustrative example. 

Consider a block of a certain size to be coded with a certain intra-coding mode (e.g., DC mode as used in HEVC). The RD cost calculation in the encoder predicts this block using DC mode, calculates the residuals, performs the transform and the quantization, determines the residual coefficients, calculates the required rate by simulating the entropy encoder, and finally performs the dequantization and the inverse transform to determine the distortion based on the reconstructed block. During this process the decoding energy estimation can be performed using the information that is present. In this case, we add the feature-specific energy for an intra predicted block $e_\mathrm{intraPU}$ to the energy required for an inverse transformation $e_\mathrm{trans}$ and for residual coefficient coding (as estimated by the amount of non-zero coefficients $n_\mathrm{coeffs}$ and their value $n_\mathrm{val}$). In accordance with the decoding energy model (\ref{eq:BFM}) the block-based energy is thus estimated by 
\begin{align}
\hat E_\mathrm{block} & =  e_{\mathrm{intraPU}} + e_{\mathrm{trans}} + n_\mathrm{coeffs}\cdot e_\mathrm{coeff}  \label{eq:blockEnergyExample} \\
 &  + n_\mathrm{val} \cdot e_\mathrm{val} \notag % + n_\mathrm{bits}.  \notag
\end{align}

Likewise, the decoding energy for the other intra as well as inter prediction modes can be estimated. 
By this means, the decoding energy can be estimated for further use in DERD-optimization for each coding mode and a certain block size.

\subsubsection{DERD Cost Calculation}
\label{secsec:codCostCalc}

For many modern encoders, RDO calculations are mainly performed on two levels (e.g., \cite{Sullivan98} for H.263, \cite{HM} for HEVC):  
\begin{itemize}
\item Precise level: On this level, the costs for the tested coding modes are calculated precisely.
Therefore, the complete residual transformation and quantization pipeline is performed. The sum of squared differences (SSD) is used to calculate the distortion. 
\item Rough level: On this level, the costs are estimated to quickly obtain useful candidates for prediction. For instance, for HEVC in the intra case, these estimates are performed for all $35$ intra modes to obtain a few (3-8) of the best modes. In the inter case, the estimates are performed to quickly find a good motion vector. The distortion costs are approximated using the Hadamard transform or the sum of absolute differences (SAD) for the predicted block. The rate is approximated solely using the bits needed to transmit the prediction mode. No residual coding or transformation is performed. As a Lagrange multiplier, as first introduced in \cite{Sullivan98}, the square root of the Lagrange multiplier used for precise calculations is employed. This is due to the SAD distortion calculations in which no squaring operation (as utilized for SSD) is performed.
\end{itemize}
In our work, we incorporate the new cost function for these two levels.  
The precise cost function reads 
\begin{equation}
\min J_\mathrm{precise} = D + {\lambda_\mathrm{R}}\cdot n_\mathrm{bits} + {\lambda_\mathrm{E}}\cdot \hat E_\mathrm{block}.  
\label{eq:CU_level_cost}
\end{equation} 

On the rough level, we compensate by multiplying with $\sqrt{5\cdot 10^6}$ to account for the fact that $\lambda_\mathrm{E}$ is situated in a different range of values (around $5\cdot 10^6$ times higher than $\lambda_\mathrm{R}$, cf. Figs. \ref{fig:lambdaQP_lR} and \ref{fig:lambdaQP_lE}) and obtain
\begin{equation}
\min J_\mathrm{rough} = D + \sqrt{\lambda_\mathrm{R}}\cdot n_\mathrm{bits} + \sqrt{5\cdot 10^6}\cdot \sqrt{\lambda_\mathrm{E}}\cdot  \hat E_\mathrm{block}.
\label{eq:PU_level_cost}
\end{equation}

\section{Evaluation}
\label{sec:eval}
\begin{figure}
\centering
\psfrag{A}[r][r]{\small{Sequence}}
\psfrag{B}[r][r]{\small{Configuration}}
\psfrag{C}[r][r]{\small{QP$_\mathrm{base}$}}
\psfrag{D}[r][r]{\small{$\Delta$QP}}
\psfrag{E}[r][r]{\small{$\tau$}}
\psfrag{F}[r][t]{\small{Specific}}
\psfrag{G}[r][b]{\small{energies $e_f$}}
\psfrag{H}[c][t]{\small{Encoder}}
\psfrag{I}[c][b]{\small{Decoder}}
\psfrag{P}[c][c]{\small{Meas.}}
\psfrag{J}[l][l]{\small{Hist(QP)}}
\psfrag{K}[l][l]{\small{PSNR}}
\psfrag{M}[l][t]{\small{Rate}}
\psfrag{N}[l][t]{\small{Estimated}}
\psfrag{O}[l][b]{\small{Energy $\hat E$}}
\psfrag{L}[c][c]{\small{Bit stream}}
\psfrag{Q}[l][l]{\small{Measured}}
\psfrag{R}[l][l]{\small{Energy}}      
\psfrag{S}[l][l]{\small{$E_\mathrm{meas}$}}
\includegraphics[width=0.35\textwidth]{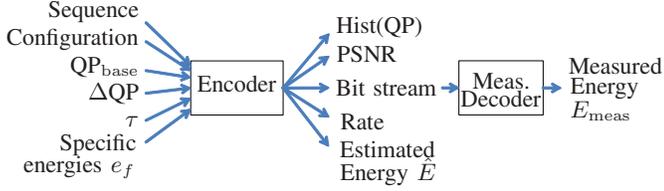}
\caption{Input and output of the modified HM-encoder. For verification, the decoding process of the output bit stream is measured (Meas.).  }
\label{fig:enc_schema}
\end{figure}

In this section, we take HEVC as a showcase standard to show that the DERDO approach can be used in practice. We evaluate the modeled decoding energies as well as the measured decoding energies. Therefore, the general test setup shown in Fig. \ref{fig:enc_schema} is used as a basis. 
As an encoder, we use a modified HM-14.0 encoder implementation \cite{HM}, which entails the functionality explained in \ref{secsec:impl}. The corresponding code is publicly available at \cite{DERDO_implementation}. As input for this encoder, we use all 8-bit sequences from the HEVC common test conditions \cite{Bossen13} and the four base QPs 22, 27, 32, 37. Furthermore, the randomaccess configuration without a frame-dependent QP offset and eleven values for the parameter $\tau$ ($0, 0.1, 0.2, ..., 1$) are used. Single and multi QP optimization is considered by setting $\Delta$QP to $0$ and $5$. The energy model we use is the simple model (FS) presented in \cite{Herglotz16b} that includes $27$ features (see Table \ref{tab:featList}). The values for the corresponding specific energies $e_f$ are trained using the method presented in \cite{Herglotz13}, 
in which the training measurements were performed for the FFmpeg decoder (the measurement method is further explained in Section \ref{secsec:meas}). Note that other models using a different amount of features or different training methods are not considered in this work because the chosen model sufficiently exhibits DERDO's potential performance.

In the following subsections, the output of the encoding process (cf. Fig. \ref{fig:enc_schema}) is investigated in detail. In Section \ref{secsec:evalLam}, we show empirical results for the $\lambda$-QP relation discussed in Section \ref{secsec:lambdaQP} by analyzing the output Hist(QP). Then, in Section  \ref{secsec:theor}, theoretical energy savings as estimated by the \textit{bit stream feature model} (BF) are discussed to show the potential performance of DERDO in terms of PSNR, rate, and the estimated decoding energy. Please note that in these two sections, if not stated otherwise, only the BQTerrace sequence is investigated to keep conciseness. The results of the other sequences are comparable. Finally, Section \ref{secsec:meas} provides results from real measurements to obtain the true decoding energy savings (Fig. \ref{fig:enc_schema}, right) for a large set of sequences.

\subsection{$\lambda$-QP Relation}
\label{secsec:evalLam}
In this section, a multi QP optimization is performed to prove that the $\lambda$-QP relation derived in Section \ref{secsec:lambdaQP} is a valid approach. 
Therefore, for every QP$_\mathrm{base}$ and every $\tau$, we allow a QP range of $\Delta\mathrm{QP}=\pm 5$ with a step size of one. The optimization is performed on the CTU level. $\lambda_\mathrm{R}$ and $\lambda_\mathrm{E}$ are initially calculated using $\rho$ and $\varepsilon$ as given in Fig. \ref{fig:lambdaQP_lR}. To ensure that QP$^*$ (which is the QP that the encoder selects most often) is located in the allowed QP range, we repeat the encoding process with adapted values for the $\lambda$s when QP$^*$ is located at the border of the QP range (the $\lambda$-values are doubled / halved which, according to (\ref{eq:QP_lambda_final}), corresponds to a QP change of approximately 3). This process is summarized in Algorithm \ref{alg:multQP}. 
\IncMargin{.5em}
\begin{algorithm}
\label{alg:multQP}
 $\rho \gets \frac{3}{2}; \varepsilon \gets \frac{3}{2}\left(c_1 \cdot p_\mathrm{I}+ c_3\right)$; 

 \For {QP$_\mathrm{base} \gets \{22, 27, 32, 37\}$ }{

\For {$\tau \gets \{0, 0.1, .., 1\}$}{

 Calculate $\lambda_\mathrm{R}, \lambda_\mathrm{E}$ using (\ref{eq:QP_lambda_final}) and (\ref{eq:tau})  with $\rho, \varepsilon, $ QP$_\mathrm{base}$, and $\tau$\;

\While {true}{

 encode\_sequence($\lambda_\mathrm{R}, \lambda_\mathrm{E}$, QP$_\mathrm{base}$, $\Delta$QP$\gets 5$)\;

QP$^*$ $\gets$ $\argmax$ Hist$($QP$)$\;

 \uIf {QP$^* == \mathrm{QP}_\mathrm{base}-\Delta \mathrm{QP}$}{

$\lambda_\mathrm{R} \gets 2\cdot \lambda_\mathrm{R}; \quad \lambda_\mathrm{E} \gets 2\cdot \lambda_\mathrm{E} $\; }

\uElseIf {QP$^* == \mathrm{QP}_\mathrm{base}+\Delta \mathrm{QP}$}{

 $\lambda_\mathrm{R} \gets 0.5\cdot \lambda_\mathrm{R}; \quad \lambda_\mathrm{E} \gets 0.5\cdot \lambda_\mathrm{E} $\;}

\Else{

 break\;
}
}
}
}

\caption{Multi QP Optimization}
\end{algorithm}
It is performed for one evaluation sequence of each class. The resulting values for QP$^*$ are saved for the validation of \eqref{eq:QP_lambda_final} and for single QP optimization. Hist$($QP$)$ is the histogram of all QPs that were selected by the encoder. 

\begin{figure}
\centering
\psfrag{000}[c][b]{$17$}
\psfrag{001}[c][b]{$22$}
\psfrag{002}[c][b]{$27$}
\psfrag{003}[c][b]{$32$}
\psfrag{004}[c][b]{$37$}
\psfrag{005}[c][b]{$42$}
\psfrag{006}[r][r]{$0$}
\psfrag{007}[r][r]{$0.2$}
\psfrag{008}[r][r]{$0.4$}
\psfrag{009}[r][r]{$0.6$}
\psfrag{010}[r][r]{$0.8$}
\psfrag{011}[r][r]{$1$}
\psfrag{012}[l][l]{\footnotesize{$\tau =1$}}
\psfrag{013}[l][l]{\footnotesize{$\tau =0.5$}}
\psfrag{014}[l][l]{\footnotesize{$\tau =0$}}
\psfrag{015}[b][t]{Hist$($QP$)$}
\psfrag{016}[t][b]{QP}
\psfrag{017}[l][l]{\footnotesize{\color[rgb]{0,0,1} QP$_\mathrm{base} = 22$}}
\psfrag{018}[l][l]{\footnotesize{\color[rgb]{0,.5,0} QP$_\mathrm{base} = 32$}}
\psfrag{019}[l][c]{\footnotesize{\color[rgb]{.749,.749,0} QP$_\mathrm{base} = 37$}}
\psfrag{020}[l][l]{\footnotesize{\color[rgb]{1,0,0} QP$_\mathrm{base} = 27$}}
\includegraphics[width=0.48\textwidth]{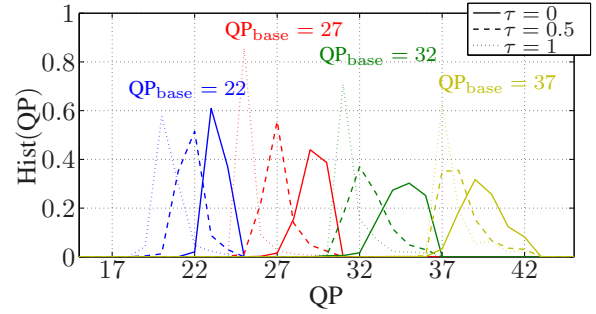}
\caption{ Histogram (relative number of occurences) of the QPs selected by the encoder for the BQTerrace sequence. The colors represent the QP$_\mathrm{base}$ values of 22, 27, 32, and 37. The solid, dashed, and dotted lines represent the $\tau$ values $0$, $0.5$, and $1$, respectively. The QP values that are mostly chosen (QP$^*$) are located at the peaks.   }
\label{fig:chosenQPs}
\end{figure}

As an example, we plot Hist$($QP$)$ for the BQTerrace sequence in Figure \ref{fig:chosenQPs} for three values of $\tau$. 
We can see that for all QPs and for all values of $\tau$ distinct peaks occur. 

\begin{figure}
\centering
\psfrag{000}[c][b]{$0.01$}
\psfrag{001}[c][b]{$1$}
\psfrag{002}[c][b]{$100$}
\psfrag{003}[c][b]{}
\psfrag{004}[c][b]{$10^6$}
\psfrag{005}[c][b]{$10^8$}
\psfrag{006}[c][b]{}
\psfrag{007}[c][b]{$10$}
\psfrag{008}[c][b]{$20$}
\psfrag{009}[c][b]{$30$}
\psfrag{010}[r][r]{$40$}
\psfrag{011}[l][l]{\ \eqref{eq:QP_lambda_final}}
\psfrag{012}[l][l]{\ QP$^*$}
\psfrag{013}[b][t]{QP}
\psfrag{014}[l][r]{$\lambda_\mathrm{E}$}
\psfrag{015}[t][c]{$\lambda_\mathrm{R}$}
\includegraphics[width=0.48\textwidth]{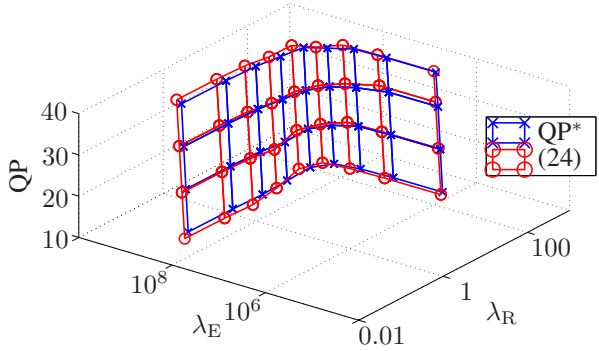}
\caption{ Experimentally derived (blue) and fitted surface (red) for the $\lambda$-QP relation (BQTerrace sequence). The corresponding $\tau$ values range from $0.1$ (right) to $0.9$ (left) in steps of $0.1$ from right to left. $\tau=0$ and $\tau=1$ are not displayed because the Lagrange multipliers would yield zero, which cannot be plotted in the logarithmic representation.  }
\label{fig:QP_lam_surf}
\end{figure}

To show that \eqref{eq:QP_lambda_final} is a valid approximation, we collect all $\{\lambda_\mathrm{R}, \lambda_\mathrm{E}, \mathrm{QP}^*\}$ triples from Algorithm \ref{alg:multQP} for the BQTerrace sequence %all sequences that were obtained 
and perform a least-squares fitting approach to calculate optimal values for $\rho$ and $\varepsilon$. The fitting algorithm is the trust region approach presented in \cite{Coleman96}. The experimentally derived surface as well as the fitted surface are shown in Fig. \ref{fig:QP_lam_surf}. 
First, the experimental result (blue) is considered. Please note that the surface is slightly tilted backwards showing that high $\lambda$s result in higher QPs. The left wing and the right wing of the surface tend to be parallel to the $\lambda_\mathrm{R}$-axis and the $\lambda_\mathrm{E}$-axis, respectively. This indicates that on these wings, the parallel axes' Lagrange multipliers only have a small influence on the optimal QP value. The fit using \eqref{eq:QP_lambda_final} (red) is very close to the experimental points. 

To evaluate the goodness of the fit, the absolute error, which is the absolute difference between QP$^*$ and the QP calculated by \eqref{eq:QP_lambda_final}, is assessed. In the case of the BQTerrace sequence, the mean absolute error of the fit is $0.73$ (maximum error $2.11$). Please note that the mean error is smaller than the minimum QP step of $1$. 
% The resulting mean absolute error (the absolute error is the absolute difference between QP$^*$ and the QP calculated by \eqref{eq:QP_lambda_final}) was found to be $0.81$ which is lower than the minimum QP step size of $1$.  
Collecting all $\{\lambda_\mathrm{R}, \lambda_\mathrm{E}, \mathrm{QP}^*\}$ triples for all sequences and calculating the least-squares fit, a mean absolute error of $0.81$ and a maximum error of $2.28$ is achieved which is only slightly higher than for the single BQTerrace sequence. This indicates that the fit is rather independent from the used input sequence. 
Hence, \eqref{eq:QP_lambda_final} is a very accurate approximation.

\subsection{Modeled Energy Savings}
 \label{secsec:theor}
In this section, the theoretical DERD-performance is analyzed. For evaluating the decoding energy, the estimated decoding energy values $\hat E_\mathrm{dec,BF}$ as returned by model BF \eqref{eq:BFM} are used. To calculate the distortion we use the YUV-PSNR \cite{Ohm12} calculated by 
\begin{equation}
D=\mathrm{PSNR}_\mathrm{YUV} = \frac{1}{8}\left( 6\cdot \mathrm{PSNR}_\mathrm{Y} + \mathrm{PSNR}_\mathrm{U} + \mathrm{PSNR}_\mathrm{V} \right) 
  \label{eq:YUV_PSNR}
\end{equation}
and the rate $R$ given by the bit stream file size. 

\begin{figure*}
\centering
\psfrag{000}[c][c]{$0$}
\psfrag{001}[c][c]{$5$}
\psfrag{002}[c][c]{$10$}
\psfrag{003}[c][c]{$15$}
\psfrag{004}[c][c]{$20$}
\psfrag{005}[r][r]{$30$}
\psfrag{006}[r][r]{$35$}
\psfrag{007}[r][r]{$40$}
\psfrag{008}[r][r]{$45$}
\psfrag{009}[c][c]{$10^5$}
\psfrag{010}[c][c]{$10^6$}
\psfrag{011}[c][c]{$10^7$}
\psfrag{012}[c][c]{$10^8$}
\psfrag{013}[r][r]{$30$}
\psfrag{014}[r][r]{$35$}
\psfrag{015}[r][r]{$40$}
\psfrag{016}[r][r]{$45$}
\psfrag{017}[b][b]{YUV-PSNR [dB]}
\psfrag{018}[b][b]{Estimated Decoding Energy [J]}
\psfrag{023}[b][b]{YUV-PSNR [dB]}
\psfrag{024}[b][b]{Bit Stream File Size [bytes]}
\psfrag{026}[c][c]{$\tau$}
\psfrag{025}[c][c]{$\tau$}
\psfrag{022}[l][l]{$\tau =0$}
\psfrag{021}[l][l]{$\tau =0.2$}
\psfrag{020}[l][l]{$\tau =0.5$}
\psfrag{019}[l][l]{$\tau =1$}
\includegraphics[width=1\textwidth]{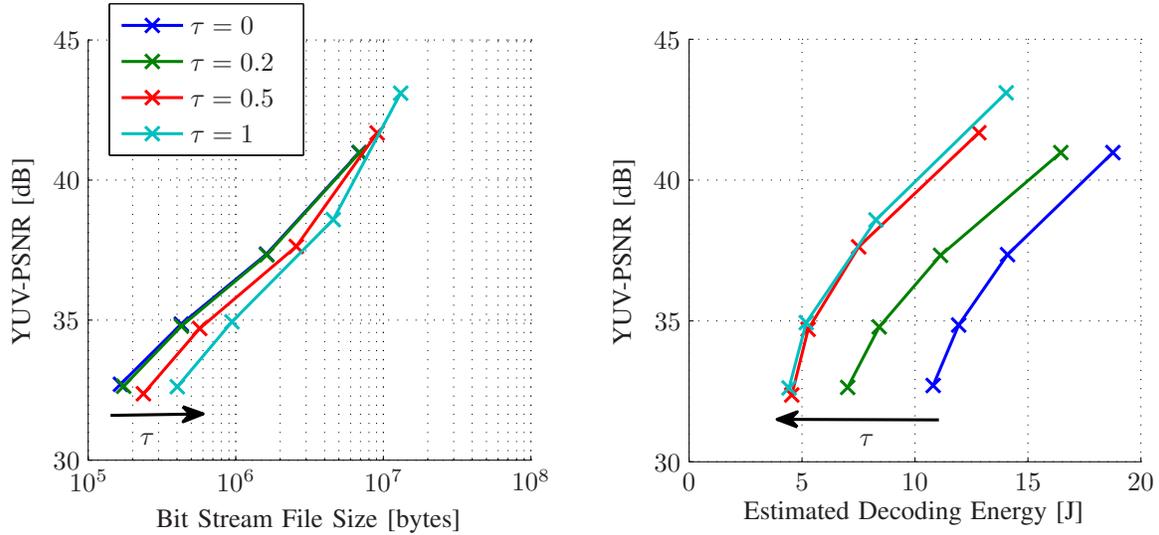}
\caption{ RD and DED curves for the BQTerrace sequence (multi QP optimization). Both y-axes show the PSNR. The x-axes show the rate in terms of the bit stream file size (left) and the estimated energy (right). Each curve corresponds to a constant value of $\tau$ and each marker denotes one base QP (QP$_\mathrm{base}=37$ bottom left to QP$_\mathrm{base}=22$ top right). The blue line corresponds to the classic RDO.   
}
\label{fig:allRD_cactus}
\end{figure*}

We base our evaluations on the rate-distortion as well as the decoding-energy-distortion curves. An example for such curves is given in Fig. \ref{fig:allRD_cactus} for the BQTerrace sequence and multi QP optimization. 
The plots show the theoretical performance of our proposed method. In the plots, each curve corresponds to a constant value of $\tau$. The left plot shows the RD performance that is commonly used in the literature. The blue line that corresponds to classic RDO ($\lambda_\mathrm{E}=0$) is located at the top left of all curves. We can see that increasing $\tau$ results in RD losses, which is caused by the increased importance of the decoding energy. 

The right plot shows the estimated decoding energies on the x-axis that correspond to the energy values that are calculated during encoding. The y-axis is the same as in the left plot. The basic RDO (the blue line) is located the furthest to the right. We can see that increasing $\tau$ now introduces energy savings as the curves tend to the left for higher values of $\tau$. This general behavior can be observed for any input sequence. However, the amount of energy savings varies depending on the input sequence and the QP. 

As interpreting these rate-distortion and energy-distortion curves is rather complicated, we calculate mean values using the well-known Bj{\o}ntegaard Delta method \cite{Bjonte01} with piece-wise cubic interpolation. For rate-distortion values, we use the Bj{\o}ntegaard Delta Rate (BDR) and for mean decoding-energy-distortion values, we replace the rate with the decoding energy in the Bj{\o}ntegaard-Delta calculus and obtain the Bj{\o}ntegaard Delta Decoding Energy (BDDE). This metric can be interpreted as the mean energy savings in percent at a constant objective visual quality. Hence, for each value of $\tau$, the two metrics BDR and BDDE describe the performance of the encoder in terms of rate and decoding energy. Please note that $\tau=0$, which is the classic RDO, is used as a reference. 

\begin{figure}
\centering
\psfrag{000}[c][b]{$0\%$}
\psfrag{001}[c][b]{$20\%$}
\psfrag{002}[c][b]{$40\%$}
\psfrag{003}[c][b]{$60\%$}
\psfrag{004}[c][b]{$80\%$}
\psfrag{005}[r][r]{$-60\%$}
\psfrag{006}[r][r]{$-40\%$}
\psfrag{007}[r][r]{$-20\%$}
\psfrag{008}[r][r]{$0\%$}
\psfrag{010}[l][l]{\small{Single QP}}
\psfrag{009}[l][l]{\small{Multi QP}}
\psfrag{011}[b][t]{BDDE}
\psfrag{012}[t][b]{BDR}
\psfrag{013}[l][l]{\footnotesize{$\tau=0$}}
\psfrag{014}[c][c]{\footnotesize{$\tau=1$}}
\psfrag{015}[l][l]{\footnotesize{$\tau=0.5$}}
\psfrag{016}[l][l]{\footnotesize{$\tau=0.3$}}
\includegraphics[width=0.48\textwidth]{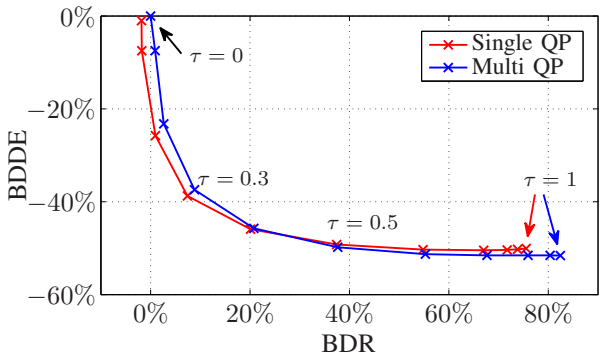}
\caption{Rate-energy diagram for comparing multi QP and single QP optimization (BQTerrace sequence with estimated decoding energies). Each marker corresponds to one value of $\tau$ in increasing order from left to right. The more the curve is shifted to the bottom left (higher energy savings and lower rate increases) the better the performance.  }
\label{fig:single_multi_BD}
\end{figure}

The BDR-BDDE-tuples can be plot as a curve as shown in Fig. %The resulting curves for the BQTerrace sequence are plotted in Fig. 
\ref{fig:single_multi_BD} (BQTerrace sequence). In the following, this plot is called rate-energy diagram. At first, we only consider the blue multi QP curve. 
Each marker on the curve corresponds to one value of $\tau$. We can see that increasing $\tau$ exhibits two effects: First, the rate increases and second, the decoding energy decreases, which was expected. Interestingly, for $\tau$ values close to $0$ and $1$, the curve shows a saturation: Close to these points, rate increases and decoding energy savings, respectively, are nearly constant. Furthermore, highest energy savings of almost $50\%$ are obtained for $\tau \approx 0.5$. 

Considering the red curve for single QP optimization (algorithm explained below) we can see that the performance is very similar and even better for small values of $\tau$, i.e. the curve is slightly shifted to the left. This shows that the single QP approach can perform even better than the multi QP approach. This can be explained by the fact that no delta QPs must be coded, thus resulting in rate savings. Note that for a fair comparison, we used the multi QP sequences coded with pure RDO ($\Delta QP=5$ and $\tau=0$) as a reference for both single QP and multi QP sequences to obtain the curves in Fig. \ref{fig:single_multi_BD}. 
Additionally, note that single QP optimization, in contrast to multi QP optimization, reflects the major advantage of a strongly reduced encoding time (on average, in our use-case, single QP optimization is ten times faster than multi QP optimization). 

Algorithm \ref{alg:singleQP} shows the single QP optimization process. 
\begin{algorithm}
\label{alg:singleQP}
Get all triples QP$^*, \lambda_\mathrm{R}, \lambda_\mathrm{E}$ from multi QP optimization\;

Discard the subset corresponding to the current sequence\;

Train $\rho$ and $\varepsilon$ using the remaining triples\;

\For {QP$_\mathrm{base} \gets \{22, 27, 32, 37\}$ }{

 \For {$\tau \gets \{0, 0.1,.., 1\}$}{

 Calculate $\lambda_\mathrm{R}, \lambda_\mathrm{E}$ using (\ref{eq:QP_lambda_final}) and (\ref{eq:tau}) 
    with $\rho, \varepsilon, $ QP$_\mathrm{base}$, and $\tau$\; 
		
 encode\_sequence($\lambda_\mathrm{R}, \lambda_\mathrm{E}$, QP$_\mathrm{base}$, $\Delta$QP$=0$)\; 
	}
}
\caption{Single QP Optimization}
\end{algorithm}
We collect all QP$^*$ and their corresponding $\lambda_\mathrm{R}$, $\lambda_\mathrm{E}$ values from multi QP optimization. Then, we use the $\{$QP$^*, \lambda_\mathrm{R}, \lambda_\mathrm{E}\}$ triples to train $\rho$ and $\varepsilon$ such that (\ref{eq:QP_lambda_final}) fits optimally in a least-squares sense (cf. training method in Section \ref{secsec:evalLam}). To prevent training on the target sequence, its triples are discarded from the training set. Afterwards, the trained values for $\rho$ and $\varepsilon$ are used for encoding with a $\Delta$QP of $0$. The resulting bit streams are used for measurements in the following.

\subsection{Measured Energy Savings}
 \label{secsec:meas}
 In order to evaluate the measured performance of the DERDO approach, we choose the Pandaboard \cite{Panda} as a decoding device. It provides a smartphone-like architecture using an ARMv7 processor for its hardware. 
As the main decoding software we select the FFmpeg framework \cite{FFmpeg} version 2.8 since it was developed for practical applications. Please note that the FFmpeg decoder makes use of dual core processing, showing that the approach is not restricted to the single core case. 
For comparison purposes, we also conduct tests using the libde265 \cite{libde} (version 0.7) and the HM-13.0 decoding software. We base our evaluation of the DERD performance on real-world energy measurements since estimated energies using the models presented in Section \ref{sec:model} may be inaccurate. To ensure that background processes do not interfere with our measurements, they are performed on runlevel 1 (where the operating system (OS) is booted with a reduced set of background processes) with disabled light emitting diodes and, as far as possible, with further supporting background services disabled. 

\begin{figure}
\centering
\psfrag{000}[c][b]{$0$}
\psfrag{001}[c][b]{$50$}
\psfrag{002}[c][b]{$100$}
\psfrag{003}[c][b]{$150$}
\psfrag{004}[c][b]{$200$}
\psfrag{005}[c][b]{$250$}
\psfrag{006}[c][b]{$300$}
\psfrag{007}[r][r]{$2.5$}
\psfrag{008}[r][r]{$3$}
\psfrag{009}[r][r]{$3.5$}
\psfrag{010}[r][r]{$4$}
\psfrag{011}[l][l]{$P_\mathrm{idle}$}
\psfrag{012}[l][l]{$P_\mathrm{dec}$}
\psfrag{013}[b][b]{\small{Power [W]}}
\psfrag{014}[t][b]{\small{Time [ms]}}
\includegraphics[width=0.48\textwidth]{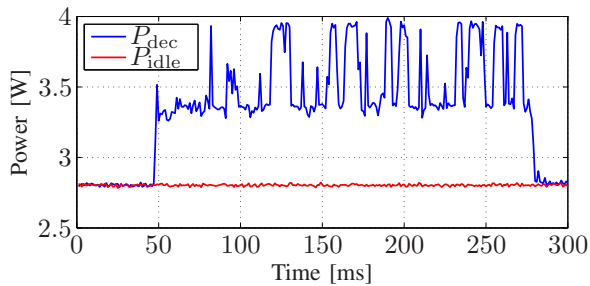}
\caption{ Power consumption in idle mode (red) and during the decoding process (blue) of the BlowingBubbles sequence. The decoding energy corresponds to the area between both curves.  }
\label{fig:decPower}
\end{figure}

The measurement setup is explained in detail in \cite{Herglotz16b}. Generally, we measure the power consumed through the main supply jack of the board such that all modules of the board are taken into account (including CPU, RAM, and periphery). A display is not attached to the device and no wireless network connection is established. As a power meter, we choose the high precision ZES Zimmer LMG95 power meter. The power consumption during the decoding process of a showcase bit stream is displayed in Fig. \ref{fig:decPower}.
The red curve shows the power consumed in idle mode (static power) which is used as a reference. The blue curve shows the power during the decoding process (static power plus dynamic power). High variations can be observed. To obtain the pure decoding energy, both curves are integrated over time and subtracted from each other. As a result, the area in between both curves represents the dynamic decoding energy which is the target of this work. 
To ensure that the measured energies are correct, multiple measurements are performed for each sequence with a subsequent significance test to ensure validity as explained in \cite{Herglotz16b}. 

\begin{figure}
\centering
\psfrag{000}[c][b]{$0\%$}
\psfrag{001}[c][b]{$20\%$}
\psfrag{002}[c][b]{$40\%$}
\psfrag{003}[c][b]{$60\%$}
\psfrag{004}[c][b]{$80\%$}
\psfrag{005}[c][b]{$100\%$}
\psfrag{006}[r][r]{$-30\%$}
\psfrag{007}[r][r]{$-20\%$}
\psfrag{008}[r][r]{$-10\%$}
\psfrag{009}[r][r]{$0\%$}
\psfrag{010}[r][r]{$0\%$}
\psfrag{011}[l][l]{\small{SlideShow (F)}}
\psfrag{012}[l][l]{\small{vidyo3 (E)}}
\psfrag{013}[l][l]{\small{BlowingBubbbles (D)}}
\psfrag{014}[l][l]{\small{BQMall (C)}}
\psfrag{015}[l][l]{\small{BQTerrace (B)}}
\psfrag{016}[l][l]{\small{Traffic (A)} }
\psfrag{017}[b][b]{BDDE}
\psfrag{018}[t][b]{BDR}
\includegraphics[width=0.48\textwidth]{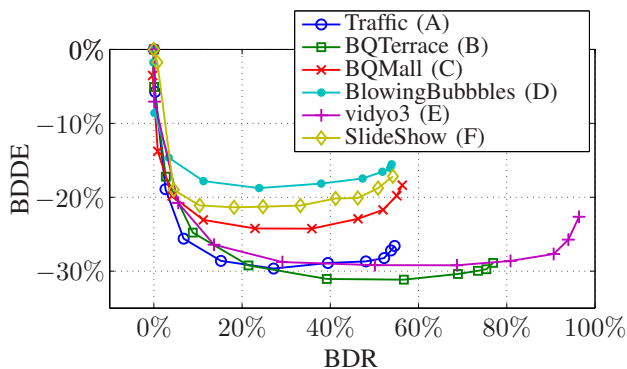}
\caption{ Rate-energy diagram for sequences from different classes of the HEVC common test conditions \cite{Bossen13} (measured energies). }
\label{fig:multiQP_BD}
\end{figure}

Figure \ref{fig:multiQP_BD} shows the rate-energy diagram with decoding energies measured for six sequences that were taken from each class of the HEVC common test conditions \cite{Bossen13}. 
First of all, we can see that the energy savings are lower than the modeled savings shown in Fig. \ref{fig:single_multi_BD}. Furthermore, the right part of the curves tend towards smaller energy savings although we had expected monotonically increasing savings. Both observations can be attributed to the energy model applied: since it is imperfect and was trained for conventionally coded bit streams, it is consequently less accurate for energy optimized bit streams. 

Additionally, we can see that highest energy savings are obtained for the BQTerrace sequence, which attains savings of more than $30\%$. Even the lowest resolution (BlowingBubbles) exhibits energy savings of nearly  $20\%$. Note that similar results can be observed for the other sequences of the corresponding class. A more detailed analysis on the energy consumption and the use of features for various rate optimized and energy optimized bit streams can be performed online at \cite{denesto}. %On this webpage, four showcase bit streams can be analyzed. The bit streams are encoded with a small and a high QP as well as a small and a high $\tau$ (BQTerrace sequence). 

\begin{figure}
\centering
\psfrag{000}[c][b]{$0\%$}
\psfrag{001}[c][b]{$20\%$}
\psfrag{002}[c][b]{$40\%$}
\psfrag{003}[c][b]{$60\%$}
\psfrag{004}[c][b]{$80\%$}
\psfrag{005}[r][r]{$-50\%$}
\psfrag{006}[r][r]{$-40\%$}
\psfrag{007}[r][r]{$-30\%$}
\psfrag{008}[r][r]{$-20\%$}
\psfrag{009}[r][r]{$-10\%$}
\psfrag{010}[r][r]{$0\%$}
\psfrag{011}[l][l]{\small{libde265}}
\psfrag{012}[l][l]{\small{HM-13.0}}
\psfrag{013}[l][l]{\small{FFmpeg}}
\psfrag{014}[b][t]{BDDE}
\psfrag{015}[t][b]{BDR}
\includegraphics[width=0.48\textwidth]{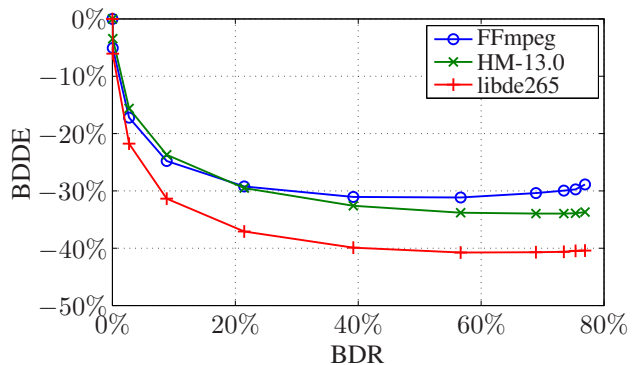}
\caption{ Rate-energy diagram for different decoder implementations (BQTerrace sequence).  }
\label{fig:BD_decoders}
\end{figure}

%\subsubsection{Decoding Software}
Finally, the proposed optimization process need not be restricted to a single software solution. To illustrate this, we measure the energy consumption of the optimized bit streams (note that optimizations were explicitly performed for the FFmpeg software) for varying software decoders, namely HM-13.0 \cite{HM} and libde265 \cite{libde}. The hardware is the same as before. The rate-energy diagram for all three decoders is displayed in Figure \ref{fig:BD_decoders} (results are shown for the BQTerrace sequence). 
We can see that energy savings are obtained for all solutions, which comes as no surprise since all decoders implement the same functionality. Surprisingly, the greatest savings are obtained for the libde265 decoder, which attains a BDDE of more than $40\%$. To explain this observation, we compare the differences between the feature numbers $n_f$ for the sequence coded with $\tau=0$ to the sequence coded with $\tau=0.5$. The feature numbers for the fractional pel interpolation and SAO show the highest differences such that, presumably, the implementation of the subpel filters and/or the SAO loop filter is relatively inefficient in the libde265 decoder software, thus resulting in higher energy savings. Note that we expect even greater savings when using a model that is explicitly trained for the libde265 decoder. 

Table \ref{tab:BDvals} summarizes the mean BDDE and BDR values for a selected subset of evaluation cases. 
We chose $\tau=0.2$ for intermediate energy savings with a small rate trade-off and $\tau=0.5$ for high energy savings. We also suggest to choose one of these values in practical applications, depending on the requirements of the user. 
\begin{table}[t]
\caption{BDR and BDDE for different test cases (single QP optimization). The values in the top $6$ rows are averaged for all sequences of the corresponding class, the remaining values hold for the BQTerrace sequence.  }
\label{tab:BDvals}
\begin{center}
\begin{tabular}{l||l||r|r||r|r}
\hline
 &                                    &  \multicolumn{2}{c||}{$\tau=0.2$}&  \multicolumn{2}{c}{$\tau=0.5$}\\
Class & Software & BDR & BDDE & BDR & BDDE \\
\hline
A & FFmpeg & $2.5\%$ & $17.7\%$ & $26.1\%$ & $27.0\%$ \\
B & FFmpeg & $3.6\%$ & $17.3\%$ & $52.3\%$ & $29.3\%$ \\
C & FFmpeg & $1.1\%$ & $13.6\%$ & $23.6\%$ & $24.5\%$ \\
D & FFmpeg & $0.4\%$ & $10.4\%$ & $20.4\%$ & $20.3\%$ \\
E & FFmpeg & $5.6\%$ & $21.4\%$ & $37.3\%$ & $30.4\%$ \\
F & FFmpeg & $2.0\%$ & $14.5\%$ & $15.2\%$ & $21.6\%$ \\
\hline
B & HM-13.0 & $2.7\%$ & $15.6\%$ & $39.2\%$ & $32.6\%$ \\ 
B & libde265 & $2.7\%$ & $21.8\%$ & $39.2\%$ & $39.9\%$ \\
\hline
\end{tabular}
\end{center}
\end{table}
We can see that approximately $15\%$ decoding energy can be saved when accepting a bitrate increase of less than $5\%$ ($\tau=0.2$). If higher rates can be tolerated, energy savings of up to $30\%$ can be achieved ($\tau=0.5$). 

Finally, the additional complexity of the encoder in terms of encoding time is assessed. As a processor, an AMD Opteron 2356 at $2.3$ GHz is used. The encoding of the BQTerrace sequence is analyzed for four QP values (QP $\in\{22, 27, 32, 37\}$) and three values of $\tau$ ($\tau\in\{0, 0.5, 1\}$). As a reference, the unchanged encoder version HM-14.0 is used with the same four QP values. The results indicate that the complexity increase is highly variable. In addition to the complexity needed for energy estimation and cost calculations, more coding modes are tested because the fast mode decisions, which are explicitly designed for regular RDO, are not optimal for the new cost function. The tests revealed that the complexity increase ranges from $17\%$ to $70\%$ and shows an average value of $42\%$. % on average, the encoder using DERDO is XX times slower than the original encoder such that complexity increases are relatively small. 

\section{Conclusions}
\label{sec:concl}
In this paper, we introduced the concept of decoding-energy-rate-distortion optimization for video coding. 
We gave an elaborate introduction into the DERD theory and derived a helpful relation between the two Lagrange multipliers and the QP. Extensive tests using various HEVC software decoders as a use case proved the performance of the proposed approach. 
The experiments showed that up to $30\%$ of the energy can be saved at the same visual objective quality when accepting bitrate increases in the order of $20-50\%$, depending on the sequence. The HEVC encoder software, including the specific energies used for the energy model and trained values for $\rho$ and $\varepsilon$, can be downloaded and freely used for personal use \cite{DERDO_implementation}. 

The proposed method of DERDO creates a vast amount of possible further research topics: Research can focus on improving the decoding energy model or the training process of the specific energies. Furthermore, in the DERDO process, the transmission energy can be considered explicitly and new encoder speed-up methods targeting the decoder energy consumption could be developed. Additionally, DERDO could be applied to other codecs such as VP9 or H.264 and even newly developed codecs could benefit from incorporating energy saving coding modes. Furthermore, potential savings with a specified hardware decoder would be of particular interest, in which the RAM modeling will have a much higher influence on the potential savings. Finally, the DERDO theory could be translated to other functionalities such as a decoding-time-rate-distortion theory that would aim at decoding time constraints for real time capability.

\bibliographystyle{IEEEtran}
% Generated by IEEEtran.bst, version: 1.14 (2015/08/26)

\begin{IEEEbiography}[{\includegraphics[width=1in,height=1.25in,clip,keepaspectratio]{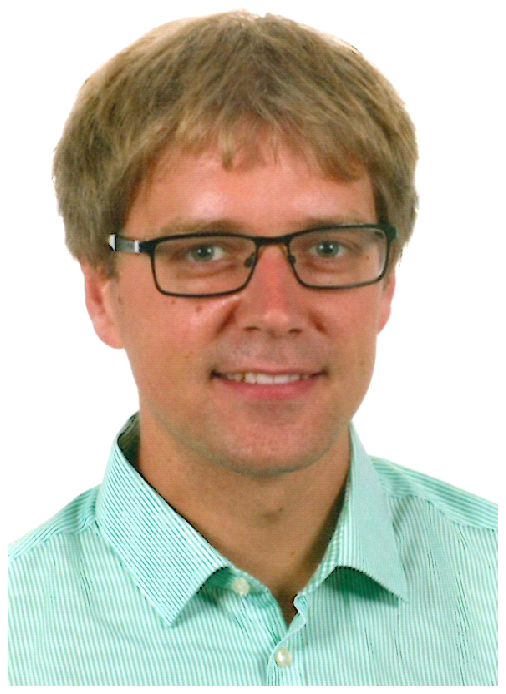}}]%
{Christian Herglotz}
received the Dipl.-Ing. in electrical engineering and information technology in 2011 and the Dipl.-Wirt. Ing. in business administration and economics in 2012, both from Rheinisch-Westf\"alische Technische Hochschule (RWTH) Aachen University, Germany. 

Since 2012 he has been a Research Scientist with the Chair of Multimedia Communications and Signal Processing, Friedrich-Alexander University Erlangen-Nürnberg (FAU), Germany. His current research interests include energy efficient video coding and fast encoding techniques. 

He received the Paul Dan Cristea Special Award 2013 for his paper "Modeling the energy consumption of HEVC intra decoding" on the 20th International Conference on Systems, Signals and Image Processing (IWSSIP). 
\end{IEEEbiography}

\begin{IEEEbiography}[{\includegraphics[width=1in,height=1.25in,clip,keepaspectratio]{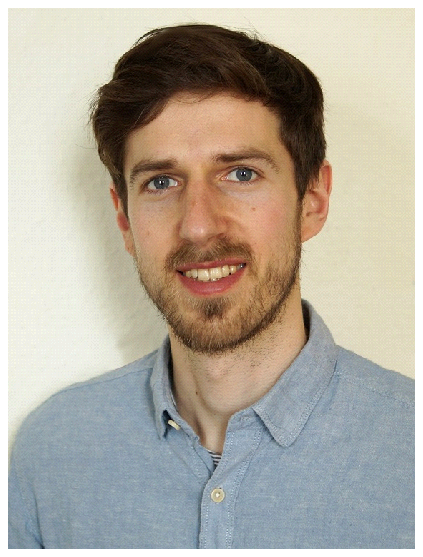}}]%
{Andreas Heindel}
received the B.Sc. degree in electrical  engineering  and the M.Sc. (Hons.) degree in systems of information and multimedia technology from the Friedrich-Alexander-Universität Erlangen–Nürnberg (FAU
), Erlangen, Germany, in 2010 and 2013, respectively. 

Since 2013, he has been a Research Scientist with
the Chair of Multimedia Communications and Signal Processing, FAU. His research interests include image  and video compression, involving different ways of scalability.

Mr. Heindel received the Fritz and Maria Hofmann Award for his master’s thesis in 2014 and the Best Tool for High Efficiency Video Coding Award from the Grand Compression Challenge at the Picture Coding Symposium in 2013.
\end{IEEEbiography}

\begin{IEEEbiography}[{\includegraphics[width=1in,height=1.25in,clip,keepaspectratio]{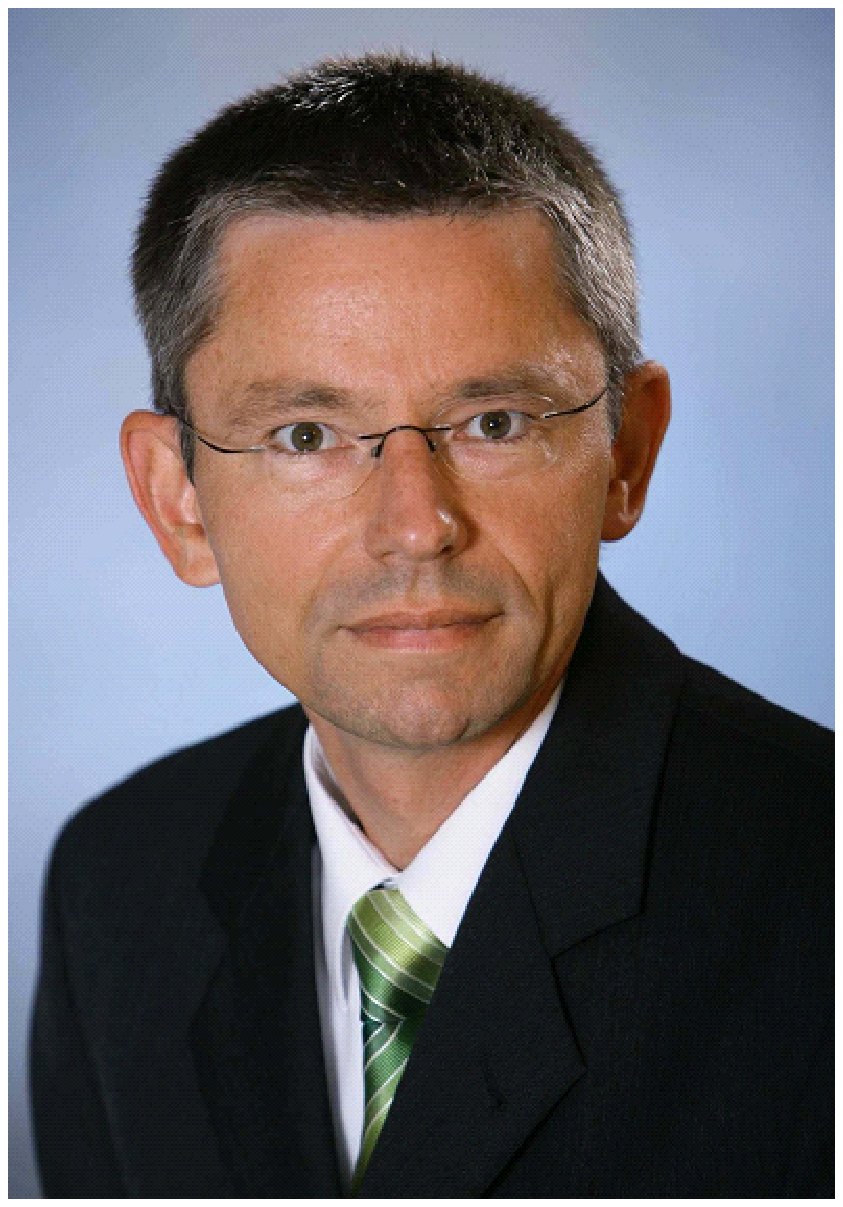}}]%
{Andr\'e Kaup}
(M'96-SM'99-F'13) received the Dipl.-Ing. and Dr.-Ing. degrees in electrical engineering from Rheinisch-Westf\"alische Technische Hochschule (RWTH) Aachen University, Aachen, Germany, in 1989 and 1995, respectively.

He was with the Institute for Communication Engineering, RWTH Aachen University, from 1989 to 1995. He joined the Networks and Multimedia Communications Department, Siemens Corporate Technology, Munich, Germany, in 1995 and became Head of the Mobile Applications and Services Group in 1999. Since 2001 he has been a Full Professor and the Head of the Chair of Multimedia Communications and Signal Processing, Friedrich-Alexander University of Erlangen-Nuremberg (FAU), Erlangen, Germany. From 1997 to 2001 he was the Head of the German MPEG delegation. From 2005 to 2007 he was a Vice Speaker of the DFG Collaborative Research Center 603. Since 2015 he serves as Head of the Department of Electrical Engineering and Vice Dean of the Faculty of Engineering. He has authored around 350 journal and conference papers and has over 70 patents granted or pending. His research interests include image and video signal processing and coding, and multimedia communication.

Dr. Kaup is a member of the IEEE Multimedia Signal Processing Technical Committee, a member of the scientific advisory board of the German VDE/ITG, and a Fellow of the IEEE. He served as an Associate Editor for IEEE TRANSACTIONS ON CIRCUITS AND SYSTEMS FOR VIDEO TECHNOLOGY and was a Guest Editor for IEEE JOURNAL OF SELECTED TOPICS IN SIGNAL PROCESSING. From 1998 to 2001 he served as an Adjunct Professor with the Technical University of Munich, Munich. He was a Siemens Inventor of the Year 1998 and received the 1999 ITG Award. 
He has received several best paper awards, including the Paul Dan Cristea Special Award from the International Conference on Systems, Signals, and Image Processing in 2013. His group won the Grand Video Compression Challenge at the Picture Coding Symposium 2013.
\end{IEEEbiography}

% that's all folks
\end{document}